\documentclass[preprint]{aastex} %preprint2 for 2-col

\usepackage{natbib}
\usepackage{graphicx}
\usepackage{epstopdf}
\usepackage{float}
\usepackage{rotating}
\usepackage{color}

\begin{document}

\slugcomment{\today}
% custom commands here
\newcommand{\bd}{{\sc{bedisk}}}
\newcommand{\br}{{\sc{beray}}}
\newcommand{\chii}{$\chi^{2}_{i}$}
\newcommand{\chis}{$\chi^{2}_{s}$}
\newcommand{\nri}{$(n,\rho_0,i)$}
\newcommand{\hnri}{$(\rho_0,n,R_{disk},i)$}
\newcommand{\gcc}{$\mathrm{g \cdot cm^{-3}} \,$}
\newcommand{\ha}{H$\alpha$}
\newcommand{\ft}{{\sc{2dDFT}}}

\title{The disk physical conditions of 48 Persei constrained by contemporaneous H$\alpha$ spectroscopy and interferometry}
\author{C.~E.~Jones\altaffilmark{1}, T.~A.~A.~Sigut\altaffilmark{1}, B.~J.~Grzenia\altaffilmark{1}, C.~Tycner\altaffilmark{2}, R.~T.~Zavala\altaffilmark{3}}
\altaffiltext{1}{Department of Physics and Astronomy, The University of Western Ontario, London, ON Canada N6A 3K7}
\altaffiltext{2}{Department of Physics, Central Michigan University, Mount Pleasant, MI 48859 USA}
\altaffiltext{3}{US Naval Observatory, Flagstaff Station, 10391 W. Naval Observatory Rd, Flagstaff, AZ 86001 USA}

\begin{abstract}

 We utilize a multi-step modelling process to produce synthetic interferometric and spectroscopic observables, which are then compared to their observed counterparts.  Our extensive set of interferometric observations of the Be star 48 Per, totaling 291 data points, were obtained at the Navy Precision Optical Interferometer from 2006 Nov 07 to 2006 Nov 23. Our models were further constrained by comparison with contemporaneous H$\alpha$ line spectroscopy obtained at the John S. Hall Telescope at Lowell Observatory on 2006 Nov 1. Theoretical spectral energy distributions, SEDs, for 48~Per were confirmed by comparison with observations over a wavelength regime of 0.4 to 60 microns from \citet{tou10} and \citet{vie17}. Our best-fitting combined model from H$\alpha$ spectroscopy, H$\alpha$ interferometry and SED fitting has a power law density fall off, $n$, of 2.3 and an initial density at the stellar surface of $\rho_0 = 1.0 \times 10^{-11}$ \gcc\ with a inclination constrained by H$\alpha$ spectroscopy and interferometry of 45$^{\circ} \pm 5^{\circ}$. The position angle for the system, measured east from north, is 121$^{\circ}$ $\pm$\ 1$^{\circ}$. Our best-fit model shows that the disk emission originates in a moderately large disk with a radius of 25 R$_{*}$ which is consistent with a disk mass of approximately 5 $\times$ 10$^{24}$ g or 3 $\times$ 10$^{-10}$ M$_{*}$. Finally, we compare our results with previous studies of 48~Per by \citet{qui97} with agreement but find, with our much larger data set, that our disk size contradicts the findings of \citet{del11}.
\end{abstract}


\section{Introduction}

Classical Be stars are distinguished by the presence of Balmer emission lines in their spectra.  As first proposed by \citet{str31}, the Balmer lines are attributed to an equatorial disk of material surrounding the star (\citealt{riv13}; \citealt{por03}). Other defining characteristics of Be stars include linearly polarized light, infrared and radio continuum excess due to radiative processes within the disk and rapid stellar rotation.  As well, these systems often exhibit variability over a range of time scales (for details, see the recent review by \citet{riv13}).

The classical B-emission (Be) star 48~Per (HD 25940, HR 1273, spectral type B3V) is well studied and located at a distance of 146 pc\footnote{based on Hipparcos parallaxes \citep{van07}}. \citet{sle49} originally classified this star as pole-on but the appearance of doubly peaked H$\alpha$ profiles reported by \citet{bur53} led \citet{ruu82} to suggest that it has an inclination of 34$^{\circ}$ to 40$^{\circ}$. Since then, the value of the inclination for this system has remained contentious. The reported changes in the spectral line shape and in brightness (see \citet{tur87} and references therein) point to periods of variability exhibited by 48 Per. However, we note that 48 Per was particularly stable over the time our observations were acquired (see the next Section for more detail).

Studies by \citet{qui97} and \citet{del11} combined interferometry with other observables for 48 Per, such as polarimetry and spectroscopy, and their work is ideally suited to detailed comparison with the results presented here. The \citet{qui97} study, hereafter `Q97', obtained interferometric observations with the Mark~III Interferometer \citep{sha88} as well as spectropolarimetric observations.  Although Mark~III has since been decommissioned, it was a predecessor to the instrument used for this study, the Navy Precision Optical Interferometer, and the two instruments share some characteristics. 48~Per was observed with six distinct interferometric baselines resulting in a set of 46 observations in the study by Q97. Through modeling they were able to place bounds on the size and inclination of the H$\alpha$ emitting region. They conclusively demonstrated, for the first time, that Be star disks could not be both geometrically and optically thick. More recently, \citet{del11} or `D11', obtained data from the Center for High Angular Resolution Astronomy (CHARA) interferometer \citep{ten05} to constrain estimates for the size of the H$\alpha$ emission region for 48~Per. Q97 confirm the nearly pole-on orientation found by \citet{sle49} by determining a minimum inclination of 27$^{\circ}$ which is also consistent with an inclination of 30$\pm 10^{\circ}$ more recently reported by D11.

The overall progression of this study is as follows; in Section \ref{sec:obs} we detail our observational program, and Section \ref{sec:model} provides an overview of the code used to calculate the theoretical disk models, along with the data pipeline we developed to analyze the model and observational data.  The results of this analysis are presented in Section \ref{sec:res}.  Finally, Section \ref{sec:sum} discusses our findings along with a comparison to other work and implications.

%%%%%%%%%%%%%%%%%%%%%%%%%%%%%%%%%%%%%%%%%%%%%%%%%
\section{Observation Program}\label{sec:obs}

\begin{figure}
\centering
\includegraphics[scale=0.6]{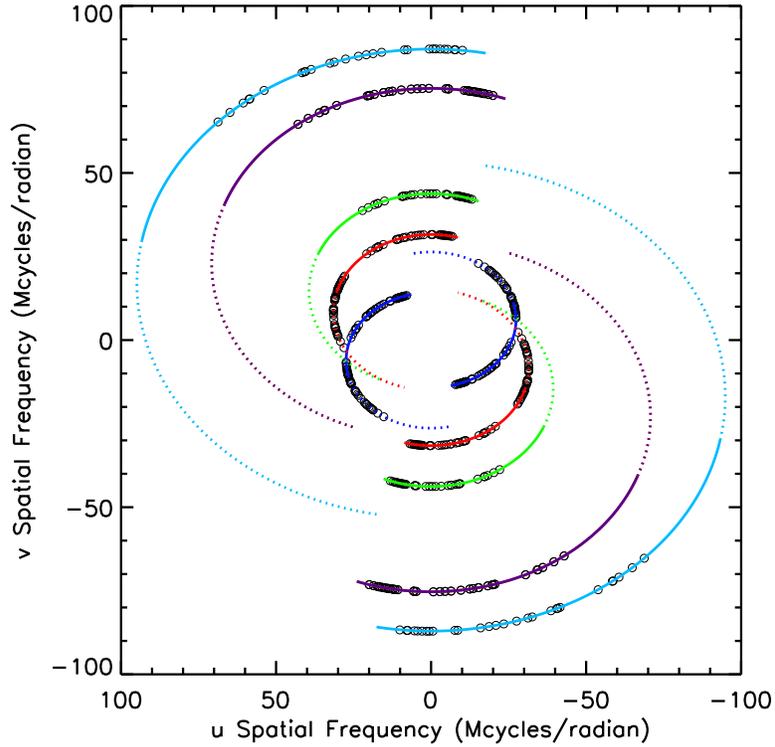}
\caption{Sky coverage in the $(u,v)$ plane for 48~Per from the H$\alpha$ containing spectral channel. The circles correspond to data points obtained at five unique baselines at the dates listed in Table~\ref{tab:npoidates2}. The arcs represent possible coverage from the meridian to 6$^{\rm h}$ east (dotted lines) and from the meridian to 6$^{\rm h}$ west (solid lines). The blue, red, green, violet and light blue colours correspond to data from baselines of 18.9, 22.2, 29.5, 51.6, and 64.4 m, respectively.}
\label{fig:uvplot}
\end{figure}

Our observations of 48~Per were obtained at the Navy Precision Optical Interferometer (NPOI), located near Flagstaff, AZ, USA.  The NPOI has an unvignetted aperture of 35 cm with an effective aperture for the observations of 12.5 cm set by the diameter of the feed system optics. See \citet{arm98} and \citet{hut16} for additional technical descriptions of this facility. Typically, observations from up to five unique baselines are obtained simultaneously; for this study, baseline lengths ranged from 18.9 to 64.4~m.  A total of 291 squared visibility measurements from a 150~\AA\ wide spectral channel containing the H$\alpha$ emission line~(i.e., centered at 6563~\AA) of 48~Per were obtained at NPOI in the autumn of 2006. The specific dates of observation are listed in Table~\ref{tab:npoidates}, and Table~\ref{tab:npoidates2} provides details such as time, $(u,v)$-space coordinates, and baseline specifier for each individual observation. Figure \ref{fig:uvplot} shows the $(u,v)$ plane coverage based on five unique baselines, where individual observations are represented by open circles and the arcs illustrate the possible coverage from the meridian to 6$^{\rm h}$ east (dotted lines) and from the meridian to 6$^{\rm h}$ west (solid lines).

\begin{figure}
\centering
\includegraphics[scale=0.6]{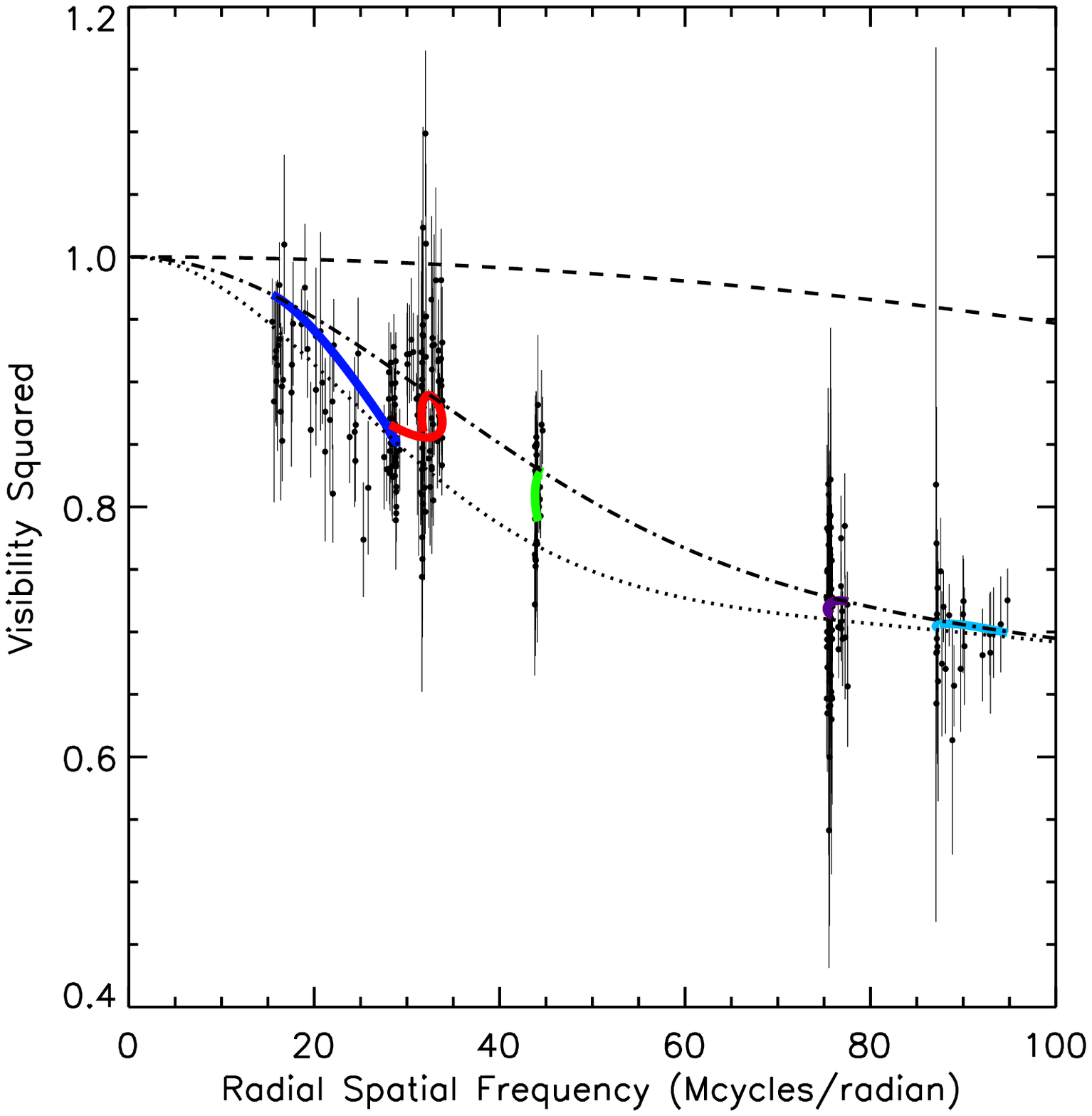}
\caption{Observed interferometric visibility squared (black dots) versus spatial frequency for 48~Per.  The dotted line represents the major axis and dashed-dotted line represents the minor axis of the disk system modeled with an elliptical Gaussian fit, GD, to the data as seen on the sky. The solid coloured curves each correspond to one specific baseline and depict changes as a function of spatial frequency resulting from diurnal motion for the same elliptical Gaussian fit as represented with the dotted lines for major and minor axes. The colours correspond to the baselines as described for Figure \ref{fig:uvplot}. The dashed line represents the central star with a uniform disk, UD, diameter of 0.306~mas.}
\label{fig:48v2obs}
\end{figure}

The squared visibilities from the H$\alpha$ channel were calibrated with respect to the continuum channels following the methodology outlined in \citet{tyc03}.  This approach allows for a more effective removal of instrumental and atmospheric effects since the H$\alpha$ and continuum channels are recorded simultaneously and correspond to the same direction on the sky (i.e., come from the same source).  Because the photosphere of the central star is not expected to be spatially resolved by our baselines at any significant level, the continuum channels are assumed to follow a uniform disk~(UD) model with an angular diameter of 0.306~mas (based on the spectral type adopted stellar parameters, including Hipparcos distance, from Table~\ref{tab:48param}).  Figure~\ref{fig:48v2obs} shows the calibrated squared visibilities from the H$\alpha$ containing channel~($V^2_{H\alpha}$) for all five unique baselines as a function of radial spatial frequency, along with a UD model curve that represents the continuum channels.  Although strictly speaking an external calibrator is not required to calibrate the interferometric squared visibilities from the H$\alpha$ channel when the continuum channels are utilized as a calibration reference, we have utilized observations of an external reference star to correct for small higher order channel-to-channel instrumental variations following the procedure described by~\citet{tyc06}.  For the purpose of these higher order corrections we utilized a nearby non-Be type star, $\epsilon$~Per(HR~1220, HD~24760), for which the observations were interleaved with those of 48~Per.

\begin{deluxetable}{cr}
\tablecolumns{2}
\tablewidth{0pc}
\tablecaption{Interferometric NPOI Observations\label{tab:npoidates}}
\tablehead{
UT Date & Number of $V^2_{H\alpha}$\\
& measurements}
\startdata
2006 Nov 07& 18\\
2006 Nov 08& 18\\
2006 Nov 09& 26\\
2006 Nov 10& 10\\
2006 Nov 11& 14\\
2006 Nov 14& 1\\
2006 Nov 15& 10\\
2006 Nov 16& 22\\
2006 Nov 17& 42\\
2006 Nov 18& 24\\
2006 Nov 20& 36\\
2006 Nov 21& 42\\
2006 Nov 22& 24\\
2006 Nov 23& 4\\
\enddata
\\[0.9ex]
\end{deluxetable}

\begin{deluxetable}{ccccc}
\tablecolumns{5}
\tablewidth{0pc}
\tablecaption{Interferometric Observations\label{tab:npoidates2}}
\tablehead{
JD$-$2,450,000 & $u$ & $v$  & $V^2_{H\alpha}$ & Baseline\tablenotemark{1}\\
&10$^6$ cycles/rad & 10$^6$ cycles/rad & $\pm$ 1 $\sigma$ &}
\startdata
4046.749 & $  17.189$ & $ -21.882$ & 0.834 $\pm$ 0.026 & AC-AE \\
4046.749 & $ -29.224$ & $   0.422$ & 0.849 $\pm$ 0.033 & AC-AW \\
4046.782 & $  21.140$ & $ -18.991$ & 0.854 $\pm$ 0.062 & AC-AE \\
4046.782 & $ -30.994$ & $  -4.121$ & 0.893 $\pm$ 0.128 & AC-AW \\
4046.815 & $  24.338$ & $ -15.402$ & 0.793 $\pm$ 0.040 & AC-AE \\
\enddata
\\[0.9ex]
\tablenotetext{1}{The baselines AC-AE, AC-AW, AW-W7, AC-W7, AE-W7 correspond to lengths of 18.9, 22.2, 29.5, 51.6, and 64.4 m, respectively.}
\parbox{0.8\textwidth}{Table~\ref{tab:npoidates2} is published in its entirety in a machine readable format.  A portion is shown here for guidance regarding its form and content.}
\end{deluxetable}

\begin{figure}
%Figure revised by Sigut 15 December 2016. Caption repeats legend, btw?
\includegraphics[scale=0.8]{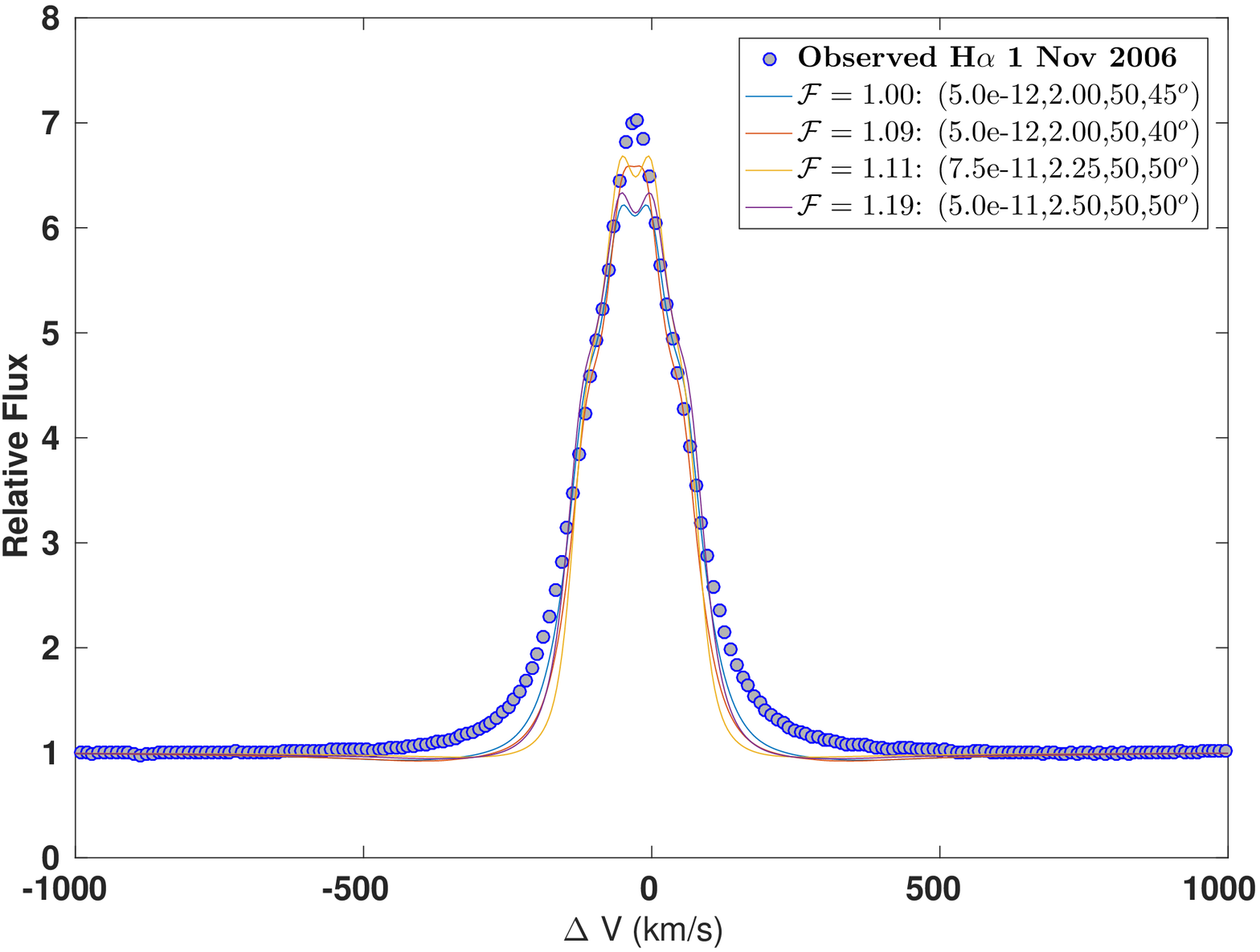}
\caption{The \ha\ line profile for 48~Per (blue circles) obtained on 2006 Nov 1 (equivalent width 28.2 \AA) is shown with a sample of our best-fitting models with \hnri = 5.0$\times 10^{-12}$ \gcc, 2.0, 50 $R_*$, 45$^{\circ}$ (blue line), \hnri\ = 5.0$\times 10^{-12}$ \gcc, 2.0, 50 $R_*$, 40$^{\circ}$ (red line), \hnri = 7.0$\times 10^{-11}$ \gcc, 2.25, 50 $R_*$, 50$^{\circ}$ (yellow line), and \hnri = 5.0$\times 10^{-11}$ \gcc, 2.50, 50 $R_*$, 50$^{\circ}$ (purple line), corresponding to ${\cal F}$/${\cal F}_{min}$ of 1.00, 1.09, 1.11 and 1.19, from top to bottom, respectively.} 
\label{fig:brackets}
\end{figure}

To complement our interferometric observations we have also obtained contemporaneous \ha\ observations with the Solar Stellar Spectrograph (SSS), an echelle spectrograph attached to the John S. Hall Telescope at Lowell Observatory (\citealt{hal94}). The emission line is singly peaked, consistent with a disk viewed more pole-on to mid inclinations. Over the course of our observing program the \ha\ emission line remained remarkably stable. Our \ha\ spectra acquired on 2006 Nov 1 and on 2006 Dec 9 that bracket the time frame of our interferometric data are indistinguishable, with \ha\ equivalent widths of 28.2 and 28.1 \AA, respectively. Based on a larger set of seven H$\alpha$ profiles obtained over the time period from 2006 Apr 10 to 2006 Dec 30, the emission profile shows only 1.4$\%$ variation based on the standard deviation of the set. The observed H$\alpha$ spectrum (blue circles) obtained on 2006 Nov 1 is shown in Figure \ref{fig:brackets} along with a sample of our best-fitting models based on our figure-of-merit value, $\cal F$, analysis discussed in the next section.

%%%%%%%%%%%%%%%%%%%%%%%%%%%%%%%%%%%%%%%%%%%%%%%%%%%%%%%%%
\section{Modelling}
\label{sec:model}
\subsection{Data Pipeline:  {\sc{Bedisk}}, {\sc{Beray}} and {\sc{2dDFT}}}\label{sub:codes}

\bd\ was developed by \citet{sig07}. It is a non-local thermodynamic equilibrium (non-LTE) modelling code which calculates self-consistent temperature distributions based on the corresponding density distribution and level populations within the disk \citep{sig07}. For the present study, the density structure within the circumstellar disk was described by a power law:
\begin{equation}
\label{eq:power}
\rho(R,Z) = \rho_0 \left(\frac{R_*}{R}\right)^n \exp \left[-\left(\frac{Z}{H}\right)^2\right]{\texttt{,}}
\end{equation}
where $\rho_0$ is the density at the disk-star boundary, $Z$ is the distance from the plane of the disk measured normal to the disk and $H$ is the vertical scale height of the disk measured perpendicular to the disk. We assume that $H$ is in approximate vertical hydrostatic equilibrium with a temperature 0.6 $\times$ T$_{eff}$ of the star, see \citet{sig09} for details.

The temperature and density distributions as well as level populations calculated with \bd\ are used as inputs to \br\ \citep{sig11}. The \bd-\br\ sequence is used to obtain a model intensity image of the disk system on the plane of the sky. \br\ calculates a formal solution of the radiative transfer equation through the disk along approximately $10^5$ rays from the observer's line of sight. The computational region extends from the photosphere to a distance (in terms of stellar radii) specified by the user. For this study, models were computed for disk sizes of 6.0, 12.5, 25.0 and 50.0 R$_*$.  

Interferometric data is in the form of a series of squared visibilities ($V^2$) as a function of spatial frequency, which is itself a representation of the interferometric baseline \citep{tho86}.  The 2D Discrete Fourier Transform (2dDFT) code \citep{sig15} takes the 2-dimensional discrete Fourier transform of the \br \ image and produces $V^2$ as a function of spatial frequency. The code then compares the model to a set of observations supplied by the user and estimates goodness-of-fit based on a reduced $\chi^2$ test. This reduction pipeline was developed by \citet{sig15} and first used to model the Be star $o$ Aquarii. In addition to determining the density distribution within the disk, our Fourier analysis is used to calculate the system's angular dimensions on the sky, the position angle of the system, the disk mass and corresponding angular momentum.

\subsection{Model Parameters}\label{sub:param}
The spectral type B3Ve was adopted for 48~Per, which is consistent across the two comparison studies of Q97 and D11, and with the Bright Star Catalog \citep{hof82}.  Further, this is in agreement with examples in the literature dating back over the past six decades (see, for example, \citet{but60} or \citet{bor60}).  The stellar parameters for the B3Ve type were determined by linear interpolation from \citealt{cox00} and are provided in Table~\ref{tab:48param}.

\begin{deluxetable}{lr}
\tablecolumns{2}
\tablewidth{0pc}
\tablecaption{Adopted Stellar Parameters for 48 Per.\label{tab:48param}}
\tablehead{Parameter & Value}
\startdata
M (M$_{\sun}$) \tablenotemark{1} & 7.6 \\
R (R$_{\sun}$) \tablenotemark{1} & 4.8 \\
L (L$_{\sun}$) \tablenotemark& 2580 \\
T$_{\rm eff} (K)$ \tablenotemark{2} & 18800 \\
$\log g$ & 4.0 \\
Distance (pc) \tablenotemark{3} & 146.2 $\pm$ 3.5 \\
Angular Diameter (mas) & 0.306 \\
\enddata
\tablenotetext{1}{adopted from Table~15.8 of \citet{cox00}}
\tablenotetext{2}{interpolated in Table~15.7 of \citet{cox00}}
\tablenotetext{3}{adopted from \citet{van07}}
\\[0.9ex]
\end{deluxetable}

\subsection{Computational Grid}\label{grid}

The parameter space was chosen to be consistent with $n$ and $\rho_0$ values that would be expected for Be star disks based on historical predictions \citep{wat86} and on contemporary studies (see, for example, section 5.1.3 of \citet{riv13} for a summary of recent results in the literature). As mentioned above, our models were computed for a range of disk sizes of 6.0, 12.0, 25.0 and 50.0 R$_{*}$. Other model parameters were varied as follows; $1.5 \le n \le 4.0$ in steps of 0.25, $1.0\times 10^{-13} \le \rho_0 \le 2.5 \times 10^{-10}$ \gcc\ in increments of 2.5 over each order of magnitude, with inclinations ranging from 20$^{\circ}$ to 65$^{\circ}$ in steps of 5$^{\circ}$. 

%%%%%%%%%%%%%%%%%%%%%%%%%%%%%%%%%%%%%%%%%%%%%%%%%%%%%%%%%%%
\section{Results}\label{sec:res}

\subsection{H$\alpha$ Spectroscopy}

Our \ha\ line profile models were compared directly to the observed spectrum obtained on 2006 Nov 1. Our model spectra were convolved with a Gaussian of FWHM of 0.656 \AA\ to match the resolving power of $10^4$ of the observed spectra. For each comparison, the percentage difference between the observed line and model prediction were averaged over the line from 6555 \AA\ to 6570 \AA\ to determine figure-of-merit value, $\cal F$, computed by,

\begin{equation}
\label{eq:fom}
    \mathcal{F}=\frac{1}{N} \sum_{i} w_i \frac{\left| F_{i}^{obs}-F_{i}^{mod} \right|}{F_{i}^{obs}},
\end{equation}
with
\begin{equation}
\label{weight}
{w_i} = \left|\frac{F^{obs}_i}{F^{obs}_c} - 1 \right|,
\end{equation}
where the sum is over all of the points in the line. $F^{obs}_i$ and $F^{mod}_i$ are the observed flux and the model flux, respectively. $F^{obs}_c$ is the observed continuum flux. Equation~\ref{weight} emphasizes the fit in the core and peak of the line while minimizing differences in the wings. Finally $\cal F$ was normalized by the best fit, i.e. ${\cal F}$/${\cal F}_{min}$, so that in our analysis the model best fit has a value of 1. This technique of matching the core of the line was found to be useful in a previous study for $o$ Aquarii \citep{sig15}. Overall, our model spectra were too weak in the wings similar to the results of D11 and \citet{sig15}. Figure~\ref{fig:brackets} shows the four best-fitting spectra within 20\% of the best-fitting model along with the observed line. The density parameters for each model are listed in the legend in the upper right of the figure along with the value of $\cal F$. The parameters in brackets in the legend correspond to $\rho_0$ in \gcc, $n$, disk size in $R_*$, and inclination angle. The parameters corresponding to our best fit are \hnri = 5.0$\times 10^{-12}$ \gcc, 2.0, 50 $R_*$, 45$^{\circ}$ (blue line on Figure~\ref{fig:brackets}). The average inclination for the four best-fit models is 46 $\pm$ 5$^{\circ}$. 

The H$\alpha$ line observed for 48~Per has an EW of 28.2 \AA\ and exhibits the singly-peaked profile we expect to see from a disk system with low to moderate inclination. Q97 estimated the lower limit for the inclination angle of 48~Per to be 27$^{\circ}$. D11 determined a best-fit inclination from their kinematic model of 30 $\pm$ 10$^{\circ}$. Our model spectra for inclinations at 30$^{\circ}$ and lower did not reproduce the observed line shape well. The model \ha\ lines were too narrow and the wings of the line were too weak. Considering a slightly larger set of 16 best-fitting models, corresponding to ${\cal F}$ within $\sim 30 \%$ of the best fit, there are 3 models with an inclination of 35$^{\circ}$, and the remainder in this set have inclinations between 40$^{\circ}$ and 55$^{\circ}$ with only one model at this highest value. The average inclination of this set is 47 $\pm$ 7$^{\circ}$. The models with the greater inclinations tended to fit the wings better since broader lines occur naturally with increasing inclination but the spectra corresponding to highest inclinations have a doubly-peaked shape unlike the observed profile. 

\begin{figure}
\centering
%Figure revised by Sigut 15 December 2016.
\includegraphics[scale=0.6]{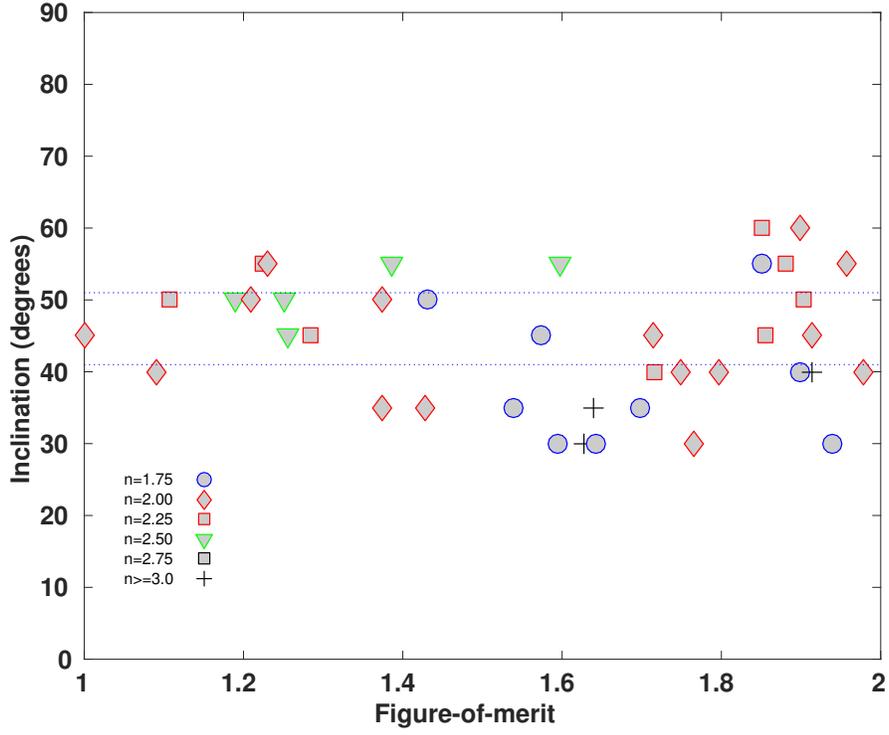}
\caption{The model inclination distribution from H$\alpha$ spectroscopy versus ${\cal F}$/${\cal F}_{min}$. The legend in the lower left shows the range of $n$ values corresponding to models for ${\cal F}$/${\cal F}_{min} \le 2$. The horizontal dotted lines correspond to the inclination $\pm 1 \sigma$ obtained from Gaussian disk GD, fits to the interferometric data.}
\label{inclination}
\end{figure}

Figure~\ref{inclination} shows the inclination versus ${\cal F}$/${\cal F}_{min}$
for all of our computed models for ${\cal F}$/${\cal F}_{min} \le 2$. The symbols in the legend in the lower left of the Figure correspond to the values of the density power law exponent, $n$. The values of the base density, $\rho_0$, vary with $n$. Generally, for small $n$, i.e. slower density fall-off with increasing distance from the central star, $\rho_0$ is also correspondingly reduced to obtain a similar amount of material in the disk to produce the H$\alpha$ emission and vice versa. The horizontal dotted lines on Figure~\ref{inclination} correspond to the inclination $\pm 1 \sigma$ obtained from Gaussian disk fits, GD, to the interferometry for ease of comparison. See Section~\ref{geofits} for more details about the geometric fits.

We see from Figure~\ref{inclination}, and as discussed above, that our best-fitting models for ${\cal F}$/${\cal F}_{min} \le 1.2$ have inclinations with 46 $\pm$ 5$^{\circ}$. However, with slight increases in the value of ${\cal F}$/${\cal F}_{min}$ to within $\sim$ 30 $\%$ we see a range in the inclination of $\sim$ 30$^{\circ}$ to 55$^{\circ}$. The lower limit of this range is consistent with the lower limit obtained by Q97 and the result of D11. Note all of our spectroscopic best-fitting models with ${\cal F}$/${\cal F}_{min} \le 2.0$ shown in Figure~\ref{inclination} corresponded to models computed with a disk size of 50 $R_*$. 

\subsection{H$\alpha$ Interferometry}

\begin{figure}
\centering
%Figure revised by Sigut 15 December 2016. Tweaked the caption also ...
\includegraphics[scale=0.6]{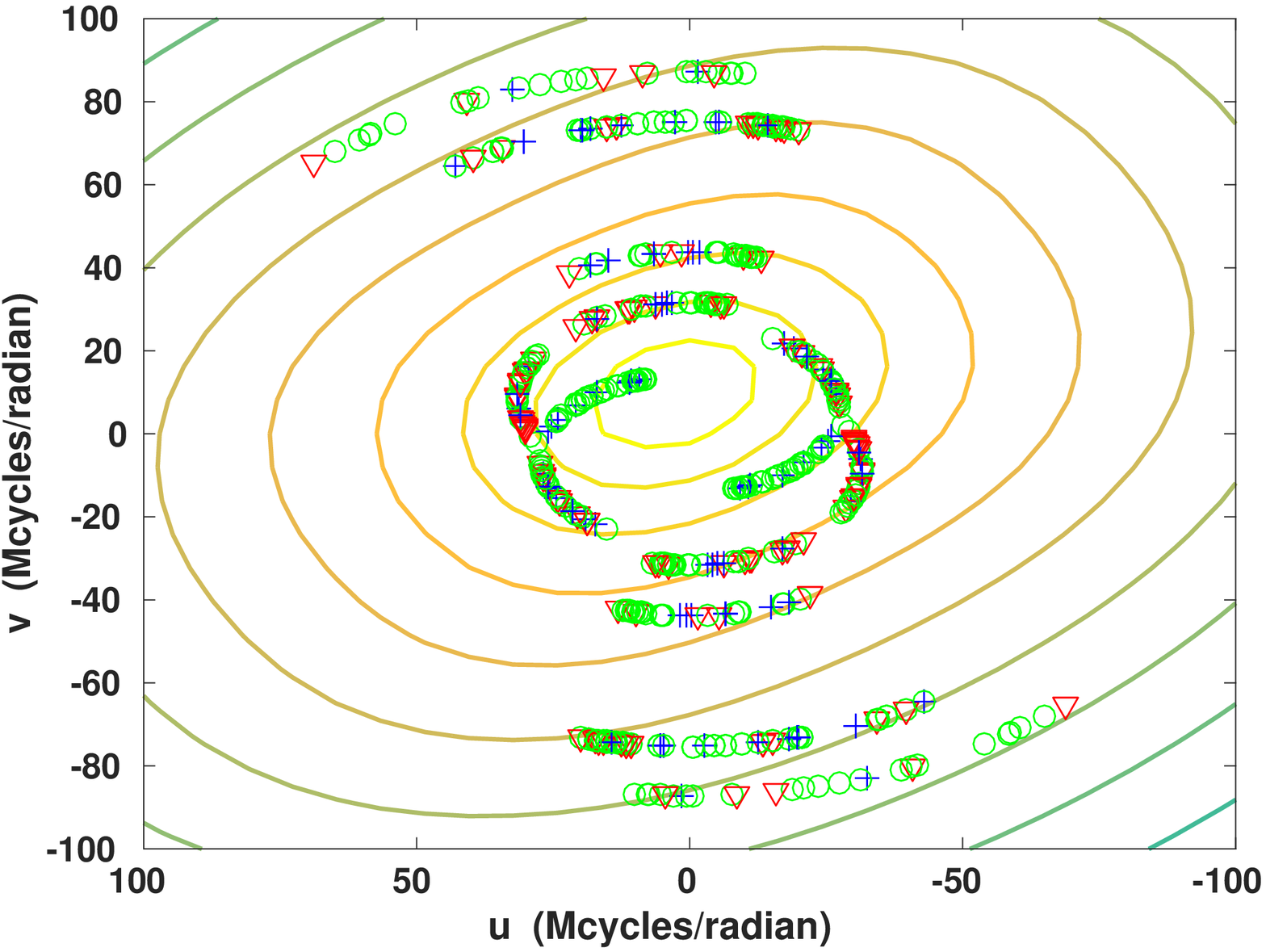}
\caption{Agreement between our best-fitting model visibilities and the NPOI observations as a function of the $(u,v)$ spatial frequencies. The model visibilities are plotted as the contours, spaced in steps of 0.05, with the innermost contour corresponding to 0.95. The best-fit model parameters are, \nri\ = 3.0, 1.0$\times 10^{-10}$ \gcc, 45$^{\circ}$. The predictions of this model at the observed spatial frequencies are plotted as green circles (181 points), red triangles (62 points) and blue plus signs (48 points). The colours represent the degree of agreement between the model and observations. The green points have $V^2$ within the errors, and the red and blue represent models that have $\chi^2$ too low and too high, respectively. The best-fit model has a reduced $\chi^2$ of 1.39 and a position angle of 121 $\pm$ 1$^{\circ}$.}
\label{best_int}
\end{figure}

The image file outputs from \br\ were fed into {\sc{2dDFT}} to obtain models of $V^2$ against spatial frequency, which were then compared directly to data obtained from interferometry by a reduced $\chi^2$ calculation. The results of the best fit to $V^2$ data, \nri\ = 3.0, 1.0$\times 10^{-10}$ \gcc, 45$^{\circ}$, are shown in Figure \ref{best_int}. The model $V^2$ symbols are plotted as green circles (181 points), red triangles (62 points) and blue plus signs (48 points). The colours represent the degree of agreement between the model visibilites and NPOI observations. The green points have $V^2$ that agree with the observed data within the errors. Given the the majority of the points are in this category (over 60$\%$) and we conclude that our model represents the data reasonably well within $\pm 1 \sigma$. The red and blue symbols represent model $V^2$ that have $\chi^2$ too low (21$\%$) and too high (16$\%$), respectively. The reduced $\chi^2$ corresponding to the best-fit model is 1.39 with a position angle, PA, of 121$\pm$1$^o$. 

\begin{figure}
\centering
%Figure revised by Sigut 15 December 2016
\includegraphics[scale=0.6]{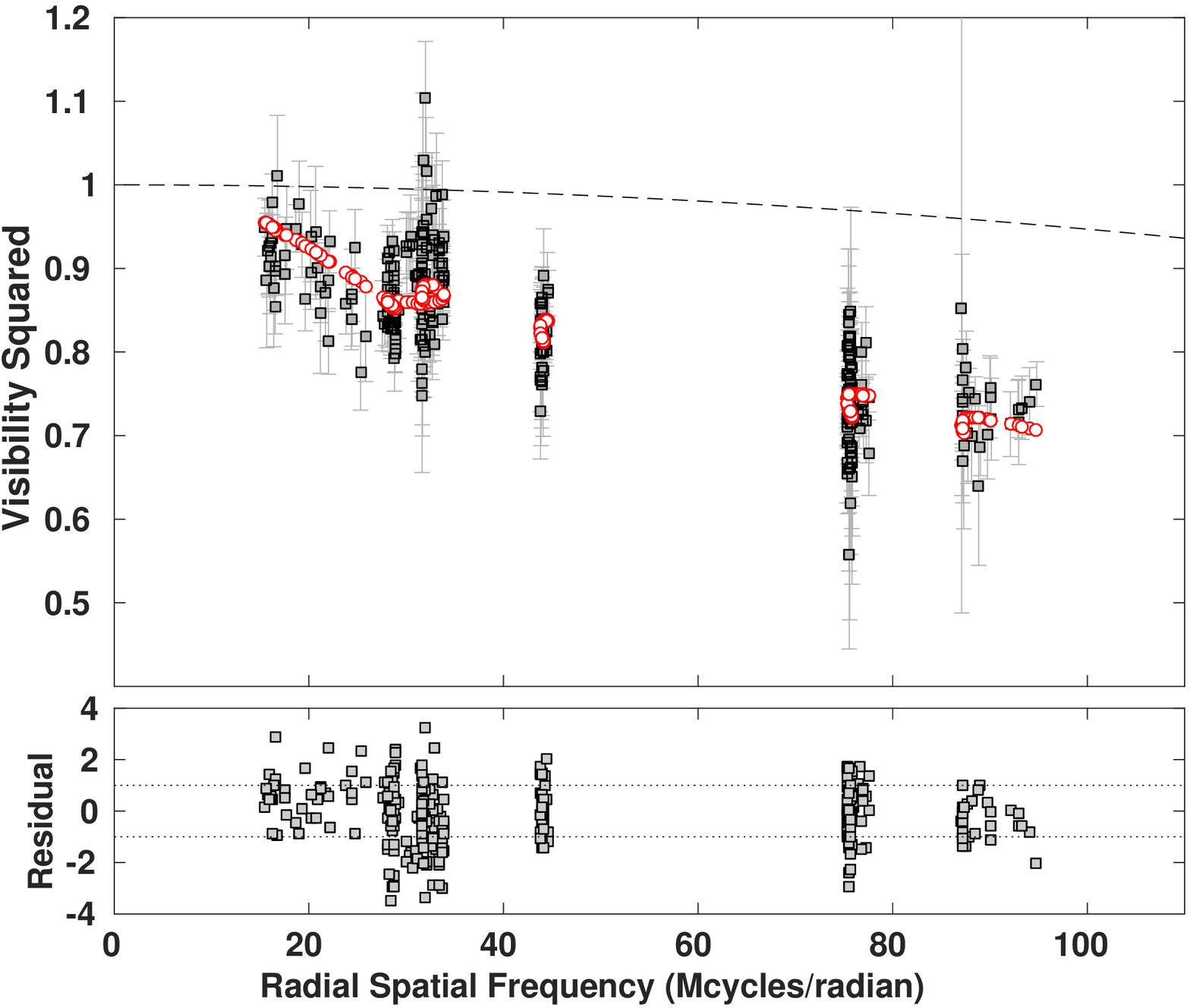}
\caption{A comparison of predicted {\sc{Beray}} visibilities to the observations, analogous to Figure~\ref{best_int}, is shown in the top panel; however, the visibilities are plotted as a function of the magnitude of the spatial frequency. The observed data, along with associated error bars $(\pm 1\,\sigma)$, are shown in black and the model in red. 
The plot is constructed using the best-fit model with the parameters listed in the caption of Figure~\ref{best_int}. 
The dashed line corresponds to the star of a uniform disk of 0.306 mas. The residuals between the model and the data are shown in the bottom panel.}
\label{modelV2}
\end{figure}

In order to assess our model predictions, we compared our predicted {\sc{Beray}} visibilities (shown in Figure~\ref{best_int}) with our best-fitting model obtained by our interferometry analysis. In Figure~\ref{modelV2}, the visibilities are plotted as a function of the spatial frequency for the same best-fit model shown in Figure~\ref{best_int}. The observed data are shown in black with the model in red. The dashed line corresponds to the star of a uniform disk of 0.306 mas. The residuals between the model and the data are shown in the bottom panel and demonstrate that our model visibilities fit the observations within $\pm 1 \sigma$. 

\begin{figure}
    \centering
    %Figure revised by Sigut 15 December 2016
    \includegraphics[scale=0.6]{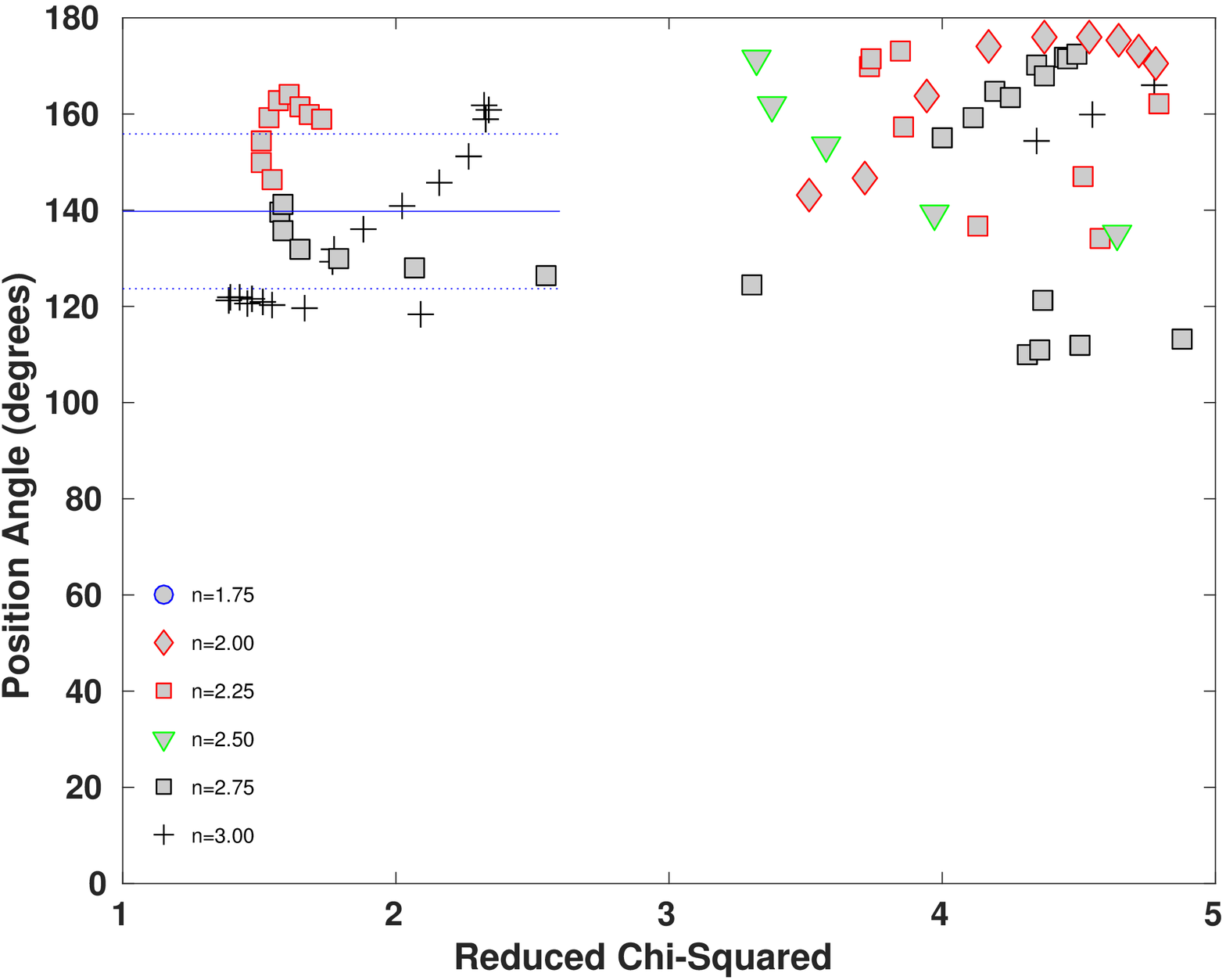}
    \caption{Plot of position angles, PA, in degrees versus reduced $\chi^2$ for $\chi^{2}/{\nu} \le 5$. The symbols in the lower left show the range of the density parameter, $n$. The horizontal lines correspond to the mean PA (solid blue line)  and the mean $\pm 1 \sigma$ (blue dotted lines) obtained from model fits to the interferometry for $\chi^{2}/{\nu} \le 2.5$.}
    \label{fig:pa_models}
\end{figure}

Figure~\ref{fig:pa_models} shows the PA for our models that correspond to $\chi^{2}/{\nu} \le 5$. Horizontal lines correspond to the mean PA (solid blue line)  and the mean $\pm 1 \sigma$ (blue dotted lines) obtained from model fits to the interferometry for $\chi^{2}/{\nu} \le 2.5$. We note that for models with $\chi^{2}/{\nu} \le 2.5$ that there is considerable scatter in the PA of about the mean of 140$^{\circ}$ of $\sim$ 15$^{\circ}$. However, the five best-fit models corresponding to $\chi^{2}/{\nu} \le 1.5$ shown on Figure~\ref{fig:pa_models} have a tight range of PAs of 121 $\pm$ 1$^{\circ}$. This is good agreement with the PA determined from the best elliptical Gaussian fit to the interferometry described next. 

\subsection{$V^2$ Geometric Fits}\label{geofits}

It is also interesting to compare the size the \ha\ emitting region and position angle of our models with geometric fits to the interferometry data. We note that geometric fits use simple functions but no physics to determine some basic physical characteristics of the disk. Here we follow the technique described in \citet{sig15} as developed by \citet{tyc06b}. Geometric fits were also computed by D11 and Q97 for 48~Per so we include a comparison with their work as well. Tables~\ref{tab:ud} and~\ref{tab:gd} compare the results of geometric fits to the visibilities using UD and GD fits, respectively. These tables also provide the axis ratios of the minor axis to major axis, the position angle of the major axis of the disk on the sky with respect to north, the fractional contribution from the photosphere of the central star to the H$\alpha$ containing interferometric signal, $c_*$, the reduced $\chi^2$ and the number of data points ($N$) used to obtain each result. The reader is referred to \citet{sig15} and references therein for more details about our geometric models.

\begin{deluxetable}{lcccccr}
\tablecolumns{7}
\tablecaption{Uniform Disk (UD) Geometric Fits based on H$\alpha$ channels.\label{tab:ud}}
\tablehead{
Study & Major Axis & Axis Ratio & Position Angle & $c_*$& $\chi^{2}/{\nu}$ & $N$\\
& [mas]&&[degrees]&[stellar&&[number\\&&&&contribution]&&of data]}
\startdata
This study & 5.70 $\pm$ 0.11 & 0.69$\pm$0.02 &119$\pm$3& 0.876$\pm$0.002 & 1.480 &291\\
D11 & 3.4 $\pm$ 0.2 & 0.77 $\pm$ 0.06 &110$\pm$ 19 &\textemdash & 0.56&3\\
\enddata

\end{deluxetable}

\begin{deluxetable}{lcccccr}
\tablecolumns{7}
\tablecaption{Elliptical Gaussian Disk Geometric Fits based on H$\alpha$ $V^2$ Data.\label{tab:gd}}
\tablehead{Study & Major Axis & Axis Ratio & Position Angle & $c_*$& $\chi^{2}/{\nu}$ & $N$\\
& [mas]&&[degrees]&[stellar&&[number\\&&&&contribution]&&of data]
}
\startdata
This study & 3.24$\pm$ 0.08 & 0.71 $\pm$ 0.03 &122$\pm$3& 0.855$\pm$ 0.003 &1.456&291\\
Q97 Modified Fit& 2.77 $\pm$ 0.56 & 0.89 $\pm$ 0.13 & 68 &0.27\tablenotemark{1}& -&46\\
D11 result & 2.1 $\pm$ 0.2 & 0.76 $\pm$ 0.05 & 115$\pm$ 33& ---&0.62&3\\
\enddata
\\[0.9ex]
\tablenotetext{1}{Not a fitted parameter.  The value is based on expected photon counts in 1~nm wide channel based on a star with the same $V$ magnitude.  For a wider spectral channel this value is expected to be closer to unity.}
\end{deluxetable}

Table~\ref{tab:ud} shows good agreement with D11 for the axis ratio and position angle for the UD fits, however, we obtain a larger major axis than D11. There may be several reasons for this discrepancy. The fitting procedure is slightly different in each study with D11 determining the disk parameters by removing the stellar contribution. More significantly, we note that we have a substantially larger set of interferometry data consisting of 291 points providing a greater sky coverage in the $(u,v)$ plane (see Figure~\ref{fig:uvplot}). Table 1 and figure 1 in D11 shows the details of their observations and $(u,v)$ plane coverage which is much less extensive compared to our data set. Q97 also fit their data for 48~Per with a UD and a ring-like model but the specific details about these geometric fits are not provided in their paper because they resulted in larger $\chi^2$ than their GD fits. However, they mention that these models were not significantly different from the results for their GD models shown in Table~\ref{tab:gd}.

A comparison of the results for the GD fits are presented in Table~\ref{tab:gd} and show good agreement with the major axis between Q97 and this study. D11 obtained a smaller major axis than we obtain however D11's result does agree with Q97 within the errors. The axis ratios point to a disk that is not viewed at large inclination angle that would result in large deviations from circular symmetry. There is agreement in the PA obtained except for the modified model (that takes the contribution of the star into account) by Q97 which gives a PA about half the other values presented in Table~\ref{tab:gd}. We also note that our definition of $c_*$ is the same as $c_p$ used by Q97, however since the filters (or spectral channels) have different widths, the values are different. Having said that, Q97 used a 1~nm wide filter and did not fit for the parameter $c_p$, but instead determined the value of this parameter based on photometric counts and the expected values based on the $V$ and $(B - V)$ index.

 Our UD and GD fits to interferometry listed in Tables~\ref{tab:ud} and~\ref{tab:gd} give predicted axis ratios of 
 0.69$\pm$0.02 and 0.71$\pm$0.03, respectively. If we assume an infinitely thin disk, these ratios translate into inclination angles of $\sim$ 46$^{\circ}$, nearly identical to our inclination of 45 $\pm$ 5$^{\circ}$ from spectroscopy. Recall the horizontal dotted lines on Figure~\ref{inclination} corresponding to the inclination $\pm 1 \sigma$ obtained from GD fits, plotted for convenience, with a range of predicted inclinations for ${\cal F}$/${\cal F}_{min} \le 2$ from our spectroscopic analysis. Clearly there is strong agreement for 48 Per's inclination obtained from geometric fits and our emission line modeling.

A comparison of Tables~\ref{tab:ud} and ~\ref{tab:gd} reveals that the major axis for the UD fits are always larger than the GD fits for each respective study. As a minor point of clarification, this is expected because the UD fits represent the major axis as the largest extent of the disk projected on the plane of the sky but for the GD fit, the size is proportional to the width of the Gaussian which only contains 68$\%$ of the light. 

Finally, we note that the best-fit elliptical Gaussian fit from our interferometry, shown in Figure~\ref{fig:48v2obs}, gives a PA of 121.65 $\pm$ 3.17$^{\circ}$ in good agreement with our PA model results shown in Figure~\ref{fig:pa_models} for our five best-fit models with PAs of 121 $\pm$ 1$^{\circ}$ corresponding to $\chi^{2}/{\nu} \le 1.5$.

\subsection{Spectral Energy Distributions}\label{SED}

Spectral energy distributions, SEDs, were also computed with {\sc beray} for wavelengths between 0.4 and 60 microns for comparison to the reported observations of \citet{tou10} (optical and near-IR) and \citet{vie17} (IR). The model SEDs were computed for the same range of density parameters and disk sizes as described in Section~\ref{grid}. The observed fluxes were de-reddened for an $E(B-V)=0.19$ \citep{dou94} following the extinction curve of \cite{Fitz99} assuming a standard $R_V$ of $3.1$. This is a relatively large amount of reddening for a nearby star like 48~Per; however, it is close to the galactic plane with a galactic latitude of just $b=-3.05^{\circ}$. We note, however, that the reddening is negligible for wavelengths greater than about 10 microns.  Using the best-fit models from H$\alpha$ and $V^2$, we found that the circumstellar contribution in the visual band to the reddening is negligible and therefore the E(B-V) must be completely of interstellar origin.

\begin{figure}
\centering
%Figure revised by Sigut 16 Mar 2017
\includegraphics[scale=0.6]{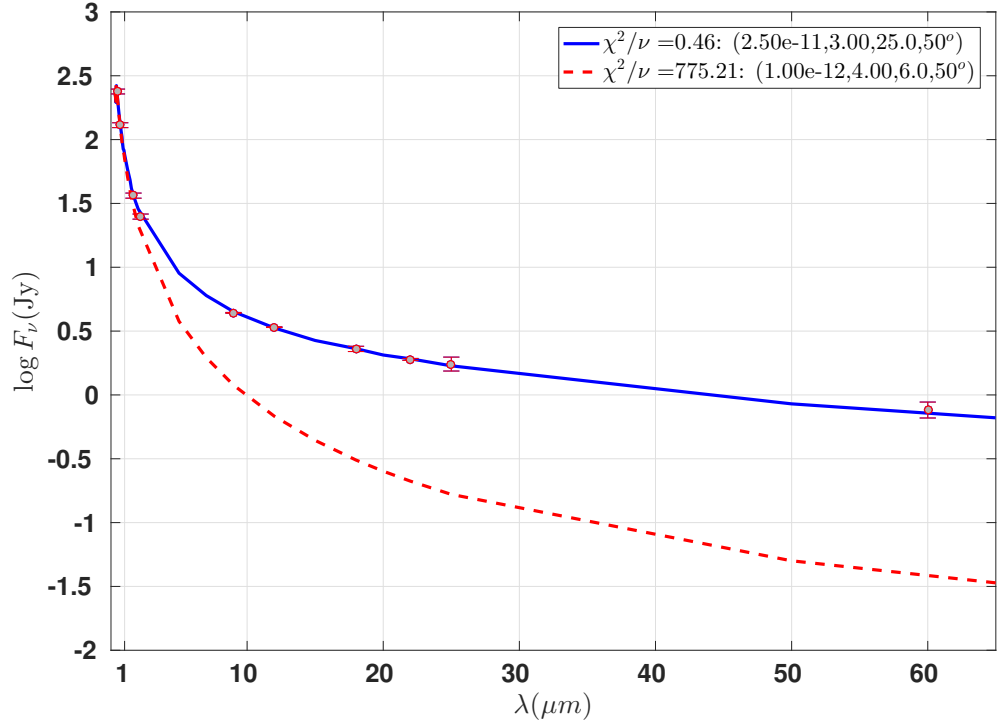}
\caption{Model SEDs for 48~Per compared with fluxes (red circles) reported by \citet{tou10} ($\lambda < 3 \mu$m) and \citet{vie17} ($\lambda > 3 \mu$m). The best-fit model SED with $\chi^2/\nu=0.46$ is shown in as the solid blue line, and an essentially disk-less model representing the stellar continuum is shown by the dashed red line. The parameters used for each model SED ($\rho_0$, n, $R_{disk}$, i) are given in the legend. }
\label{fig:sed44}
\end{figure}

Figure~\ref{fig:sed44} shows the best-fit model corresponding to a $\chi^{2}/{\nu}=0.46$ with parameters \hnri\ = 2.5$\times 10^{-11}$ \gcc, 3.0, 25 $R_*$, and $i=50^{\circ}$. We adopted an absolute error of $\pm 0.02$ in the $\log$ for the fluxes of \cite{tou10} and used the reported errors of \cite{vie17} for the longer wavelengths. A second model computed with parameters simulating an essentially disk-less system, is also shown, illustrating the underlying stellar continuum and the magnitude of the IR excess due to the disk. The best-fit SED shown in Figure~\ref{fig:sed44} has a slightly steeper power-law index compared to the H$\alpha$ fit, although it is consistent with power-law of the best-fit model to the interferometric visibilities. However, there are a number of additional models with $\chi^2/\nu$ of the order of unity, and this point is further discussed in the next section. Finally, while the best-fit model SED inclination of $i=50^{\circ}$ is consistent with the previous H$\alpha$ and $V^2$ fits, we note that the SED is a poor constraint on disk inclination at infrared wavelengths \citep{wat87}.

\subsection{Combined Results from Spectroscopy, Interferometry, and SED fits}\label{sec:combined}

Table~\ref{tab:results1} summarizes our model best-fit results based on H$\alpha$ spectroscopy, H$\alpha$ interferometry and SED fits. We note that the best models from spectroscopy, interferometry and SED fitting are reasonably consistent. H$\alpha$ spectroscopy and H$\alpha$ interferometry are consistently best-fit with models of a disk size of 50 R$_{*}$ while the best-fit SED corresponds to a disk size of 25 R$_{*}$ in our grid. Although we note that the set of best-fitting SEDs corresponding to $\chi^2/\nu\leq 1$ span a range of disk size from 6 to 50 $R_{disk}$.  

\begin{figure}
\centering
%Figure revised by Sigut March 16, 2017
\includegraphics[scale=0.6]{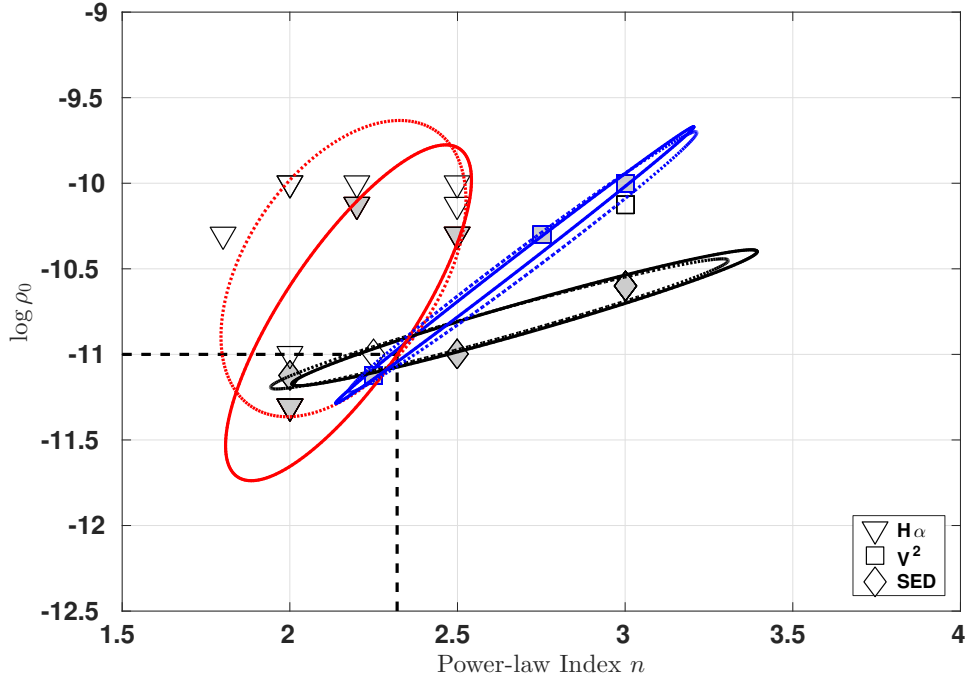}
\caption{Summary of the best-fitting disk density parameters, $n$ and $\rho_0$, for 48~Per. The solid red ellipse encloses the models that fit the H$\alpha$ line profile within 20\% of the best-fit (filled triangles), whereas the dotted red ellipse encloses the models that fit within 50\% (open triangles). The blue ellipse encloses models that fit the interferometric $V^2$ within 20\% of the best fit (solid squares, all with $\chi^2/\nu\leq 1.67$) while the dotted ellipse encloses the models that fit within 50\% (open squares with $\chi^2/\nu\leq 2.0$). The solid black ellipse encloses models that fit the optical/IR SED with $\chi^2/\nu\leq 1$, wheres the dotted ellipse encloses the models with $\chi^2/\nu\leq 1.4$. The overlap region suggests that a model with $n=2.3$ and $\log(\rho_o)=-11.0$ (indicated by the dotted black lines) is the one most consistent with all three diagnostics.}
\label{summary}
\end{figure}

Figure~\ref{summary} summarizes the best-fitting models for H$\alpha$ spectroscopy, H$\alpha$ interferometric $V^2$ and SED fitting. The solid ellipses and filled symbols represent models within 20$\%$ of the best-fit from the H$\alpha$ line profile (red ellipses, triangles) and H$\alpha$ interferometric $V^2$ (blue ellipses, squares) and for $\chi^2/\nu\leq 1$ from the SED fitting (black ellipses, diamonds). The dotted ellipses and unfilled symbols, using the same colours and symbols for each diagnostic as before, represent models within 50$\%$ of the best-fit from the H$\alpha$ line profile and H$\alpha$ interferometric $V^2$ and for $\chi^2/\nu\leq 1.4$ from the SED fitting and demonstrate the robustness of our fitting procedure.

Some of the symbols presented on Figure~\ref{summary} represent more than one model since the same value of $\rho_0$ and $n$ may have been selected with different inclination and $R_{disk}$. For the H$\alpha$ spectroscopy there are four models within 20$\%$. The parameters corresponding to these models are shown in Figure~\ref{fig:brackets}. For the H$\alpha$ interferometry there are 19 models within 20$\%$ corresponding to $\chi^{2}/{\nu}$ from 1.39 to 1.67. These 19 models have an average $n = 2.7 \pm 0.3$, $\rho_0 = (5.5\pm4.2) \times 10^{-11}$ \gcc, and inclination of $38^{\circ}$ $\pm$\ 12$^{\circ}$. There are 5 models corresponding to $\chi^2/\nu\leq 1$ from the SED fitting. These models have a range of $n$ from 2.0 to 3.0 while $\rho_0$ ranges from $7.5 \times 10^{-12}$ \gcc to $2.5 \times 10^{-11}$ \gcc. As mentioned above these models span a range of $R_{disk}$ sizes and interestingly all have an inclination of $50^{\circ}$ with the exception of one model with an inclination of $30^{\circ}$. We emphasize as discussed in Section~\ref{SED} that the SED is not a good constraint on inclination.

\begin{deluxetable}{lccc}
\tablecolumns{4}
\tablewidth{0pc}
\tablecaption{Best-fit H$\alpha$, V$^2$ and SED model results.\label{tab:results1}}
\tablehead{
Fit & n & $\rho_0$ [g cm$^{-3}$] & $i$ [degrees]  }
\startdata
H$\alpha$ Profile & 2.0 & $5.0 \times 10^{-12}$& $45 \pm 5$\\
$V^2$ & 3.0 & $1.0 \times 10^{-10}$ & $45 \pm 12$\\
SED& 3.0& $2.5 \times 10^{-11}$ & ($50$) \tablenotemark{1}\\
\enddata
\tablenotetext{1}{The inclination is not well constrained by the SED.}
\\[0.9ex]
\end{deluxetable}

%
% The new results are :
%
% R90 = 24.8 = 25
% M_gm = 4.5834e+24
% M/M* = 3.0319e-10
% J_gmcm2s = 2.9955e+44
% J/J* = 4.1327e-08
%
% + no inner region for SED.
%
% As based on n=2.32, rho0=1.0e-11 as in Figure 9. I have updated the numbers below but not altered the text.
% That's it for me 
%

The dashed lines in Figure~\ref{summary} show the model ($n = 2.3$ and $\log \rho_0 = -11.0$) corresponding to the intersection of H$\alpha$ spectroscopy, H$\alpha$ visibilities and optical/IR SED fitting that is most consistent with these three observational data sets. As discussed above the model fits to the H$\alpha$ spectroscopy and interferometry 
corresponded to disk sizes of 50~R$_{*}$. By taking our best-fit image computed with \br\ at $i = 0^{\circ}$ (face-on) and constrained by H$\alpha$ interferometry, we can integrate from the central star along a radial rays over distance to obtain a better estimate of the extent of the H$\alpha$ emitting region. At large distances from central star, the H$\alpha$ emission tends to originate from an increasingly diffuse disk. Therefore, we chose to integrate until 90$\%$ of the H$\alpha$ flux is contained within the disk. Our best-fit model with $n = 2.3$ and $\log \rho_0 = -11.0$, corresponding to spectroscopy, the visibilities and SED fitting as shown by the dashed lines in Figure~\ref{summary} gives us R$_{90}$/R$_{*}$ of 25, where R$_{90}$ represents the radial distance corresponding to 90$\%$ of the H$\alpha$ emission. We note that this calculation includes more of the H$\alpha$ flux compared to what is enclosed within the extent of the major axis of the geometric GD (defined as FWHM of the Gaussian) listed in Table~\ref{tab:gd}. This integrated disk size corresponds to a mass for the H$\alpha$ emitting region of 5 $\times$ 10$^{24}$ g or 3 $\times$ 10$^{-10}$ M$_{*}$. We note that this is likely a lower limit to the disk mass in the H$\alpha$ emitting region since as mentioned above, our models produce H$\alpha$ spectra that are too weak in the wings. 

Finally, following the prescription in \citet{sig15}, we use our disk mass to determine the angular momentum, J$_{\rm{disk}}$, in the disk compared to the central star's momentum, J$_*$. We determined the critical velocity consistent with the stellar parameters listed in Table~\ref{tab:48param} and assuming 80$\%$ critical rotation this gives an equatorial velocity of 360 km s$^{-1}$ for this calculation. For the model corresponding to the best-fit from  all three diagnostics, with a disk mass of 5 $\times$ 10$^{24}$ g or 3 $\times$ 10$^{-10}$ M$_{*}$, we obtain a value for J$_{{\rm disk}}/$J$_{*}$ of 4 $\times$ 10$^{-8}$.

%%%%%%%%%%%%%%%%%%%%%%%%%%%%%%%%%%%%%%%%%%%%%%
\section{Discussion and Summary}\label{sec:sum}

Our best-fit models corresponding to our H$\alpha$ spectroscopy, interferometry, and SED fits are summarized in Table~\ref{tab:results1}. Figure~\ref{summary} is a graphical representation of the best-fitting regions corresponding to the constraints based on different observational data sets. Table~\ref{tab:results1} and Figure~\ref{summary} show that 48~Per has a moderately dense disk with values of $n$ $\sim$ 2 to 3 and $\log$ $\rho_0$ $\sim$ -11.7 to -9.6 or $\rho_0$ $\sim$ 2.0 $\times 10^{-12}$ to 2.5 $\times 10^{-10}$ \gcc.
The radial extent with all of the best-fitting models for the H$\alpha$ and V$^2$ fits correspond to the largest disks (50 R$_*$) in our grid. The model fits for smaller disk sizes in our computational grid resulted in poorer fits for the H$\alpha$ spectroscopy and $V^2$. The combined results from all three diagnostics overlap for a model with $n = 2.3$ and $\rho_0 = 1.0 \times 10^{-11}$  \gcc\ corresponding to the dashed lines in Figure~\ref{summary}.

From spectroscopic analysis, there are four models within 20$\%$ of ${\cal F}$/${\cal F}_{min}$ value, and 19 models within 20$\%$ with $\chi^{2}/{\nu}$ ranging from 1.39 to 1.68 corresponding to the visibilities. These models point to a disk inclination of of 45$^{\circ} \pm 5^{\circ}$. From the SED fitting there are five models with $\chi^2/\nu\leq 1$. These models span a range of $i$ $\sim$ 30$^{\circ}$ to 50$^{\circ}$. This is consistent with early studies on Be star disks using infrared continuum measurements that demonstrated, especially for low to moderate inclinations, that viewing angle is not well constrained by these measurements \citep{wat87}.

As discussed in Section~\ref{sec:res} our model spectral lines were too weak in the wings. Therefore, for line fitting purposes we used a core-weighted formula for our figure of merit, $\cal F$, which places more emphasis on the central portion of the line (recall Figure \ref{fig:brackets}). D11 also found that it was not possible to fit the broad wings in the H$\alpha$ line for 48 Per. They adopted an ad hoc scheme to account for non-coherent electron scattering, a process which redistributes absorbed line photons resulting in broader lines. This process has been well studied in the literature (see, for example, \citet{mih78} for a detailed treatment) but it is difficult to properly account for in models because it lacks an analytic solution. The fact that our models were weak in the wings may also be due to this process. Alternatively, the poorer fit in the wings could also be due to the fact that a single value of $n$ for each model was adopted for this study. Finally, as briefly discussed in Section~\ref{sec:combined}, we calculated the critical velocity of the star directly from the stellar parameters listed in Table~\ref{tab:48param}. If the stellar parameters resulted in an under-estimate of the disk rotation, then potentially this could contribute to the fact that our model spectra were too narrow in the wings.

As reported above in Section~\ref{sec:res} the best-fitting models for H$\alpha$ spectroscopy and interferometry from our grid corresponded to 50 R$_*$. However, as previously discussed, by taking our best-fit model with parameters with $n = 2.3$ and $\log \rho_0 = -11.0$, obtained from interferometry, spectroscopy  and SED fitting (see the dashed line on Figure~\ref{summary}), we calculated a better approximation for the radial extent of the H$\alpha$ emitting region of R$_{90}$/R$_{*} =$ 25. As mentioned in Section~\ref{sec:combined} this model dependent calculation includes more H$\alpha$ flux than a geometric fit would contain. Consequently, the calculated disk size is correspondingly bigger than some results in the literature. Other studies have determined the radial extent of the H$\alpha$ emitting region for sets of Be stars and specific stars. For example, using H$\alpha$ spectroscopy \citet{han86} find the emitting regions of 20 R$_*$ for 24 bright Be stars, \citet{sle92} determine sizes of 7 to 19  for a sample of 41 Be stars, and more recently \citet{arc16} find a concentration of disk sizes of 5 to 20 R$_*$. Size estimates based on interferometric GD models encompassing 80\%\ of a star's brightness at FWHM for H$\alpha$ emitting regions for 12 Be stars are shown in \citet{riv13}.  Estimated radii range from 3.24 R$_*$ for $\beta$ CMi \citep{tyc05} to 16.36 R$_*$ for $\psi$~Per (D11). Our model result of 25 R$_*$ is greater, as expected, than some of the sample averages presented above but is in general agreement with other results in the literature.

Finally, based on our estimated size of the emitting region of 48~Per we determine the mass and angular momentum content of the disk. For the model corresponding to the best fit from all three diagnostics we obtain a disk mass of 5 $\times$ 10$^{24}$ g or 3 $\times$ 10$^{-10}$ M$_{*}$ and J$_{{\rm disk}}/$J$_{*}$ of 4 $\times$ 10$^{-8}$. This value for the disk mass represents a lower limit since our models were too weak in the wings but is in reasonable agreement with the result presented by \citet{ste03} for 48 Per of 11.2 $\times$ 10$^{-10}$ M$_{\odot}$ or 2.23 $\times$ 10$^{24}$ g. Using a similar technique as the work presented here, \citet{sig15}, found the mass and angular momentum of disk of the late spectral type Be star, $o$~Aqr, of 1.8 $\times$ 10$^{-10}$ M$_{*}$ and 1.6 $\times$ 10$^{-8}$ J$_{*}$, respectively. 

\citet{jon11} analyzed variability in the \ha\ equivalent widths for a sample of 49 Be stars.  They determined that over the time frame of their study, which overlaps our observations, 48~Per was remarkably stable. Given the apparent stability of this system, it is interesting to compare our best-fit values of $n = 2.3$ and log $\rho_0$ of -11.0 with other values presented in the literature. \citet{wat87} modeled 48 Per using a power law fall-off for the density distribution constrained with data from the Infrared Astronomical Satellite (IRAS). Their model is described in terms of an opening angle and stellar parameters are slightly different from our work, but the results are similar. \citet{wat87} obtain an $n$ of 2.5 and a range of log $\rho_0$ of -11.8 to -11.5.  \citet{vie15} studied the continuum emission of this system using pseudo-photosphere model and obtained $n$ of 2.5 and log $\rho_0$ of $\sim -11.48$. Most recently, \citet{vie17} used IRAS, Japanese Aerospace infrared satellite (AKARI) and Wide-field Infrared Survey Explorer (WISE) observations to constrain their viscous decretion disk model. \citet{vie17} find $n$ of 2.9, 2.8 and 2.7 and log $\rho_0$ of -11.4, -11.4 and -11.5, using IRAS, AKARI and WISE data, respectively. Despite the differences in the models and adopted stellar parameters, the values obtained for 48 Per with \citet{wat87}, \citet{vie15} and \citet{vie17} are remarkably similar. \citet{vie17} report that $n \leq 3.0$ indicates a dissipating disk. During the time of our observations, the 48 Per is quite stable. However, \citet{vie17} mention that disk dissipation seems to occur over much longer time scales so it would be interesting to follow this system over a more extended time frame. In future, we plan to extend our modeling technique to include a two-component power law for the density parameter, $n$, to account for changes in disk density with radial distance from the star.

\acknowledgements
 The authors would like to thank an anonymous referee for comments, suggestions and questions that helped to improve this paper. BJG extends thanks to Dave Stock, Anah\'{i} Granada, and Andy Pon for their mentorship as well as their helpful feedback on prior versions of this article.  BJG also recognizes support from The University of Western Ontario. CEJ and TAAS wish to acknowledge support though the Natural Sciences and Engineering Research Council of Canada. CT acknowledges support from Central Michigan University and the National Science Foundation through grant AST-1614983, and would like to thank Bryan Demapan for assistance with the interferometric data processing. The Navy Precision Optical Interferometer is a joint project of the Naval Research Laboratory and the US Naval Observatory, in cooperation with Lowell Observatory and is funded by the Office of Naval Research and the Oceanographer of the Navy. We thank the Lowell Observatory for the telescope time used to obtain the H$\alpha$ line spectra, and the US Naval Observatory for the NPOI data that were presented in this work. This research has made use of the SIMBAD database, operated at CDS, Strasbourg, France.  

\clearpage

\pagebreak
	
\bibliography{ms.bib}

\end{document}